\documentclass[%
 reprint,
 amsmath,amssymb,
 12pt,
 aps,
 onecolumn
]{revtex4-2}

\usepackage{graphicx}% Include figure files
\usepackage{dcolumn}% Align table columns on decimal point
\usepackage{bm}% bold math
\usepackage{bbold}
\usepackage{physics}
\usepackage{comment}
\usepackage{xcolor}
\usepackage{hyperref}
\begin{document}

%\preprint{APS/123-QED}

\title{Qutrit-based Synthetic Three-Level System}

\author{Surajit Sen}
\thanks{Corresponding author}
\email{ssen55@yahoo.com}
% \altaffiliation[]{Centre of Advanced Studies and Innovation Lab\\
% 18/27 Kali Mohan Road, Tarapur, Silchar 788003, India}%Lines break automatically or can be forced with \\
\author{Tushar Kanti Dey}%
 \email{tkdey54@gmail.com}
\affiliation{%
 Centre of Advanced Studies and Innovation Lab\\
 18/27 Kali Mohan Road, Tarapur, Silchar 788003, India %\textbackslash\textbackslash
}%


%\affiliation{
%\affiliation{
%\affiliation{%


\date{\today}% It is always \today, today,

\begin{abstract}
We present a theoretical framework based on the $SU(3)$ group to construct synthetic three-level configurations from a two-qutrit system consisting of two three-level subsystems. Utilizing its underlying algebraic structure and a set of nine $SU(3)$ entangled states, we show that the system Hamiltonian can be mapped onto an effective synthetic three-level manifold without introducing Rydberg states. We investigate the entanglement dynamics of these synthetic configurations by introducing the $SU(3)$ I-concurrence and a generalized Wootters-type $SU(3)$ concurrence as quantitative measures of entanglement in such system.
\end{abstract}



\maketitle

%\tableofcontents

\section{Introduction}
\par
In quantum information science, atomic, computational, and entangled states constitute distinct yet fundamental components that collectively underpin the architecture of modern quantum technology. Atomic states refer to the intrinsic energy eigen-states of physical quantum system, such as the ground and excited states of an atom, which are indeed the fundamental building blocks of quantum hardware. Computational basis states arise from the mapping of these physical atomic states onto logical qubits, typically denoted as $|0\rangle, |1\rangle, |00\rangle, |01\rangle$, etc., enabling the encoding and manipulation of quantum information in a standardized framework. Entangled states, by contrast, embody non-classical correlations between two or more qubits, where the joint quantum state cannot be factorized into individual subsystem states. Thus, while atomic states provide the physical substrate and computational basis states serve as information carriers, the engineered entangled states encode the complex quantum correlations essential for implementing advanced quantum protocols. 
\par 
Recent experimental advances have enabled the controlled generation of entangled qubit pairs which exhibit rich light-matter dynamics in highly tunable Rydberg atomic platforms. Prominent Rydberg implementations include optical tweezer arrays for high-fidelity gates~\cite{Saffman2010,Levine2019}, optical lattices and quantum gas microscopes for many-body Hubbard physics~\cite{Browaeys2020}, ultra-cold ensembles in magneto-optical traps for Rydberg-EIT and nonlinear optics~\cite{Pritchard2010}, and trapped-ion platforms~\cite{Feldker2015,Higgins2017,Zhang2020} etc. Across these platforms, the van der Waals interaction between Rydberg states play a pivotal role in enabling the formation of desired collective states which includes entangled states. 
The dynamics in the presence of entanglement can be most effectively understood within the framework of engineered three‑level system.
\par
In a standard three-level electromagnetically induced transparency (EIT) system, resonant probe and coupling fields create bright and dark states as coherent superpositions of bare atomic states ~\cite{Harris97,Fleischhauer2005,Sen2015}. In contrast, Rydberg-EIT operates in a distinctly collective regime within the composite Hilbert space $\mathcal{H}^{2\otimes 2} = \mathcal{H}_c^{2} \otimes \mathcal{H}_t^{2}$, where both two-qubit entanglement and interatomic van der Waals interaction facilitate an effective three-level structure \cite{Mohapatra2007, Pritchard2010, Walker2012, Pritchard2013, Firstenberg2016, Saffman2016, Adams2019, Ziemkiewicz2020, Finkelstein2022}. To illustrate this mapping in the control-target architecture, we consider a two-qubit system, each consisting of a ground state $|g\rangle$ and an excited state $|e\rangle$ forming the manifold $\{|g_c g_t\rangle, |\Phi_{ge}^+\rangle\}$ where, where $|g_c g_t⟩$ denotes the collective ground state and $|\Phi_{ge}^+\rangle = \frac{1}{\sqrt{2}}(|g_c e_t\rangle + |e_c g_t\rangle)$ is the singly-excited symmetric Bell state \cite{Moller2008}. Within this framework, the double-excitation state $|e_c e_t\rangle$ is energetically suppressed, while the antisymmetric counterpart $|\Psi_{ge}^-\rangle$ remains decoupled because the interaction Hamiltonian is symmetric under particle exchange.
In this two-qubit scenario, the introduction of the Rydberg state $|r\rangle$ as third level is indispensable for constructing the synthetic collective EIT basis $\{|g_c g_t\rangle, |\Phi_{ge}^+\rangle, |\Phi_{rg}^+\rangle\}$, where the states $|e_c e_t\rangle$ and $|\Phi_{er}^{\pm}\rangle$ involve excitations outside the target EIT manifold, together with the doubly excited Rydberg state $|r_c r_t\rangle$ due to strong interaction-induced Rydberg blockade mechanism. Thus, in an engineered three-level system such as the Rydberg EIT manifold, two entangled states emerge from two distinct manifolds associated with the excited and Rydberg states. 
\par
Despite significant progress in Rydberg-atom based architecture, the primary limitation of such system arises from the short-lived lifetime of high-lying Rydberg states, which affects the achievable coherence time and overall gate fidelity \cite{Saffman2010,Walker2012}. Furthermore, the steep $1/R^6$ distance dependence of the van der Waals interaction imposes rigorous constraints on the effective blockade radius, thereby limiting both the system’s scalability and the robustness of the entangling operations against positional fluctuations \cite{Urban2009, Wilk2010,Zeybek2023}. These challenges necessitate the investigation of alternative architectures that circumvent the van der Waals interactions while facilitating high-fidelity state engineering without compromising the purity of the entangled state. 
\par
To move beyond the Rydberg-mediated architecture involving Bell states, we propose an alternative entanglement framework composed of two three-level subsystems forming a two-qutrit system. The extension from two-level to three-level quantum systems is nontrivial, as such systems are conventionally classified into three distinct configurations, namely, $V$ (Vee), $\Xi$ (Cascade), and $\Lambda$(Lambda)-type  configurations, depending on their respective transition selection rules \cite{Yoo1985, Sen2012}. Recent studies have shown that the $SU(3)$ group provides a natural framework for describing these configurations through distinct pairs of $SU(3)$ shift operators \cite{Nath2008,Sen2012}. To capture the entanglement scenario of two-qutrit systems under this symmetry group, we recently introduced a class of $SU(3)$ entangled states in the composite Hilbert space $\mathcal{H}^{3\otimes 3} = \mathcal{H}_c^{3} \otimes \mathcal{H}_t^{3}$ together with an inequality that generalizes the $SU(2)$ qubit formalism to $SU(3)$ qutrit system ~\cite{Sen2024}. Furthermore, these states were shown to provide a robust framework for high-dimensional quantum teleportation protocols ~\cite{Sen2025} extending Bennett's original proposal~\cite{Bennett1993}. 
\par
In this paper, we develop a theoretical framework for constructing synthetic three-level systems from the entanglement structure of a two-qutrit manifold based on the recently introduced $SU(3)$ entangled states \cite{Sen2024}. Unlike conventional two-qubit Bell-state implementations that rely on van der Waals interactions, the present formulation offers a framework based on a synthetic three-level configurations without invoking such interactions. To quantify entanglement in such systems, we formulate the $SU(3)$ I-concurrence together with a generalized Wootters concurrence \cite{Wootters1998,Rungta2001}.
\par
The remainder of this paper is organized as follows. In Section II, we delineate the algebraic structure of the $SU(3)$ group within the control-target framework and demonstrate how this symmetry gives rise to three distinct Hamiltonian in the composite Hilbert space $\mathcal{H}^{3 \otimes 3}$. Based on the $SU(3)$ entangled states, Section~III identifies the collective states within the relevant manifold that participate in the dynamics and serve as the building blocks for constructing the model Hamiltonians of the different synthetic configurations. In Section IV, we formulate the I-concurrence and the generalized SU(3) Wootters concurrence for the synthetic three-level systems. Finally, in Section V, we summarize our results and discuss the future outlook. 

\section{The Models in composite Hilbert space}

\par 
The two-qutrit system resides in the composite Hilbert space $\mathcal{H}^{3\otimes 3} = \mathrm{span}\{|{3}_c\rangle\} \otimes \mathrm{span}\{|{3}_t\rangle\}$, where each subsystem is spanned in the local basis $|3_x\rangle$ where $x \in \{c,t\}$. The composite $SU(3)$ operators acting on this space are defined as $O^{(c)} = O \otimes \mathbb{I}_3$ and $O^{(t)} = \mathbb{I}_3 \otimes O$ where $O$ denotes the $SU(3)$ shift operator. In the computational basis,
\begin{align}
|1\rangle &= \begin{pmatrix}1 \\ 0 \\ 0\end{pmatrix}, \quad
|2\rangle = \begin{pmatrix}0 \\ 1 \\ 0\end{pmatrix}, \quad
|3\rangle = \begin{pmatrix}0 \\ 0 \\ 1\end{pmatrix},
\label{eq1}
\end{align}
the off-diagonal $SU(3)$ shift operators responsible for transitions $|j\rangle \leftrightarrow |k\rangle$, ($j \neq k$) are given by $T_+ = |1\rangle\langle2|$, $V_+ = |1\rangle\langle3|$, $U_+ = |2\rangle\langle3|$ along with their hermitian conjugate operator with corresponding the diagonal projection operator $P_j = |j\rangle\langle j|$. In this framework, the algebraic structure of the composite operators $S_{ab}^{(x)} = |a_x\rangle\langle b_x|$ and $P_i^{(x)} = |i_x\rangle\langle i_x|$ is given by, 
\begin{subequations}
\begin{align}
[S_{ab}^{(x)}, S_{cd}^{(x')}] &= \delta^{xx'} \left( \delta_{bc} S_{ad}^{(x)} - \delta_{da} S_{cb}^{(x)} \right), \\
[P_i^{(x)}, S_{jk}^{(x')}] &= \delta^{xx'} (\delta_{ij} - \delta_{ik}) S_{jk}^{(x)}, \\
\{S_{ab}^{(x)}, S_{cd}^{(x')}\} &= \delta^{xx'} \left( \delta_{bc} S_{ad}^{(x)} + \delta_{da} S_{cb}^{(x)} \right) + 2(1-\delta^{xx'}) S_{ab}^{(x)} S_{cd}^{(x')}, \\
\{P_i^{(x)}, S_{jk}^{(x')}\} &= \delta^{xx'} (\delta_{ij} + \delta_{ik}) S_{jk}^{(x)} + 2(1-\delta^{xx'}) P_i^{(x)} S_{jk}^{(x')},
%\end{aligned}
\end{align}
Furthermore, introducing shift vectors, $S_+^{(x)} = S_{jk}^{(x)}$, $S_-^{(x)} = S_{kj}^{(x)}$ and $S_3^{(x)} = \frac{1}{2}(P_j^{(x)} - P_k^{(x)})$ along with $[O_i^{(c)}, O_j^{(t)}] = 0$ for all $O_i, O_j \in \{T, U, V, P\}$,
we have following relations, 
\begin{align}
%\begin{aligned}
[S_3^{(x)}, S_{\pm}^{(x)}] &= \pm S_{\pm}^{(x)}, & [S_+^{(x)}, S_-^{(x)}] &= 2S_3^{(x)}, \\
\{S_3^{(x)}, S_{\pm}^{(x)}\} &= 0, & \{S_+^{(x)}, S_-^{(x)}\} &= (P_j^{(x)} + P_k^{(x)}), \\
\{S_3^{(x)}, S_3^{(x)}\} &= \tfrac{1}{2}(P_j^{(x)} + P_k^{(x)}), & \{S_{\pm}^{(x)}, S_{\pm}^{(x)}\} &= 0. 
\end{align}
\label{eq2}
\end{subequations} 
In contrast to conventional three-level systems where the $V$-, $\Xi$-, and $\Lambda$-type configurations are described by the pairs of $SU(3)$ shift operators $\{V,T\}$, $\{T,U\}$, and $\{U,V\}$ \cite{Nath2008,Sen2012}, the Hilbert space $\mathcal{H}^{3\otimes 3}$ admits three distinct synthetic configurations constructed from the composite shift operators $\{V^{(x)},T^{(x)},U^{(x)}\}$ and generate the $VT$, $TU$, and $UV$ configurations within a collective entangled-state basis. The total Hamiltonian of two-qutrit system is given by
\begin{align}
H^\text{M} = H^\text{M}_{\text{0}} + H^\text{M}_\text{I}, \quad M \in \{\, VT\,, TU\,, UV\, \}. 
\label{eq3}
\end{align}
For the $VT$ configuration, the free and interaction parts of the Hamiltonian are given by, 
\begin{subequations}
\begin{align}
H^\text{VT}_{\text{0}} &=\frac{1}{3} \sum_{x=c,t}\Big[E_{11}^\mathrm{VT}P_{1}^{(x)}+ E_{22}^\mathrm{VT} P_{2}^{(x)} + E_{33}^\mathrm{VT}P_{3}^{(x)}\Big], \\
\label{eq4a}
H^\mathrm{VT}_{\text{I}} &= \sum_{x=c,t}\Big[\frac{\Omega_{13}^{(x)}}{2}\big(V_+^{(x)} + V_-^{(x)}\big) + \frac{\Omega_{12}^{(x)}}{2}\big(T_+^{(x)}+T_-^{(x)}\big)\Big],
%\label{eq4b}
\end{align}
\label{eq4}
\end{subequations}
where $\Omega_{ij}^{(x)}$ is the Rabi coupling strength and $E_{ii}^{\rm VT}$ denotes the diagonal energy terms, which are functions of the detuning offset $\Delta_{ij} = \omega_{ij} - \omega_{L,ij}$. Here $\omega_{ij} \equiv \omega_i - \omega_j$ is the bare energy gap between the $i-$ and $j-$th levels, and $\omega_{L,ij}$ is the frequency of the driving field. Proceeding similar way we have, 
\begin{subequations}
\begin{align}
H^\text{TU}_{\text{0}} &= \frac{1}{3} \sum_{x=c,t}\Big[E_{11}^\mathrm{TU}P_{1}^{(x)}+ E_{11}^\mathrm{TU}P_{2}^{(x)} + E_{33}^\mathrm{TU}P_{3}^{(x)}\Big], \\
H^\text{TU}_{\text{I}} &= \sum_{x=c,t}\Big[\frac{\Omega_{12}^{(x)}}{2}\big(T_+^{(x)}+T_-^{(x)}\big)+\frac{\Omega_{23}^{(x)}}{2}\big(U_+^{(x)} + U_-^{(x)}\big)\Big],
%+ U_{\text{onsite}}(\cdot) + V_{\text{exch}}(\cdot) 
\end{align}
\label{eq5}
\end{subequations}
for the $TU$-configuration and, 
\begin{subequations}
\begin{align}
H^\text{UV}_{\text{0}} &= \frac{1}{3}\sum_{x=c,t}\Big[E_{11}^\mathrm{UV}P_{1}^{(x)}+ E_{22}^\mathrm{UV}P_{2}^{(x)} + E_{33}^\mathrm{UV}P_{3}^{(x)}\Big], \\
H^\text{UV}_{\text{I}} &= \sum_{x=c,t}\Big[\frac{\Omega_{23}^{(x)}}{2}\big(U_+^{(x)}+U_-^{(x)}+\frac{\Omega_{13}^{(x)}}{2}\big(V_+^{(x)} + V_-^{(x)}\big)\Big], 
%+ U_{\text{onsite}}(\cdot) + V_{\text{exch}}(\cdot) 
\end{align}
\label{eq6}
\end{subequations}
for the $UV$-configuration, respectively with corresponding diagonal eigen energies, detuning term and the coupling constants. 
\par 
To reveal the entanglement structure of a nonlocal two-qutrit system, we evaluate the commutator between the free and interaction Hamiltonian from Eqs.~(\ref{eq4}-\ref{eq6}). Using Eq.~(\ref{eq2}), we find the following  nonvanishing commutation relations, 
\begin{subequations}
\begin{align}
[H^\text{VT}_{0}, H^\text{VT}_{I}] = \frac{1}{6} \sum_{x=c,t} \left[ \Omega_{13}^{(x)} \Delta E_{13}^\mathrm{VT} \left(V_+^{(x)} - V_-^{(x)}\right) + \Omega_{12}^{(x)} \Delta E_{12}^\mathrm{VT} \left(T_+^{(x)} - T_-^{(x)}\right) \right],
\end{align}
for the $VT$-configuration,
\begin{align}
[H^{\text{TU}}_{0}, H^{\text{TU}}_{I}] = \frac{1}{6} \sum_{x=c,t} \left[ \Omega_{12}^{(x)} \Delta E_{12}^\mathrm{TU} \left(T_+^{(x)} - T_-^{(x)}\right) + \Omega_{23}^{(x)} \Delta E_{23}^\mathrm{TU} \left(U_+^{(x)} - U_-^{(x)}\right) \right],
\end{align}
for the $TU$-configuration,
\begin{align}
[H^{\text{UV}}_{0}, H^{\text{UV}}_{I}] = \frac{1}{6} \sum_{x=c,t} \left[ 
\Omega_{23}^{(x)} \Delta E_{23}^\mathrm{UV} \left(U_+^{(x)} - U_-^{(x)}\right) 
+\Omega_{13}^{(x)} \Delta E_{13}^\mathrm{UV} \left(V_+^{(x)} - V_-^{(x)}\right)\right],
\end{align}
\label{eq7}
\end{subequations}
for the $UV$-configuration, respectively with $\Delta E_{ij}^\mathrm{M}=E_{ii}^\mathrm{M}-E_{jj}^\mathrm{M} \neq{0}$. We note that the non-vanishing commutators between the free and interaction Hamiltonian arise not only from local coherence, but also from the coherent superposition of multiple transition pathways across two entangled qutrit sites.

\section{Synthetic Three-level Systems}
\par 
To construct the synthetic three-level system with its constituent states residing in a single manifold, it is necessary to identify the bright states, defined by the condition
$\langle g_c,g_t \vert H_{\mathrm{I},c-t}^\mathrm{MN} \vert \psi_{ab} \rangle \neq 0$,
where $\ket{g_c g_t}$ is the separable collective state and $\ket{\psi_{ab}}$ is an entangled state. 
The complete set of $SU(3)$ computational basis states for two-qutrit system obtained from Eq.~(\ref{eq2}) is
\begin{align}
\mathcal{C}^{(2)} = \big\{ 
|1_c 1_t\rangle, |1_c 2_t\rangle, |1_c 3_t\rangle, |2_c 1_t\rangle, |2_c 2_t\rangle, |2_c 3_t\rangle, |3_c 1_t\rangle, |3_c 2_t\rangle, |3_c 3_t\rangle 
\big\}. 
\label{eq8}
\end{align}
Here, the states $|1_c1_t\rangle$, $|2_c2_t\rangle$ and $|3_c3_t\rangle$ are separable states, which we refer to as reference or intermediary states for the $VT$, $TU$ and $UV$ configurations, respectively. To identify the bright states we require knowledge of the full spectrum of the two-qutrit entangled states which dictate the underlying entanglement dynamics of the composite system. In a recent study, using the representation theory of $SU(3)$ group, we developed the complete spectrum of $SU(3)$ entangled states \cite{Sen2024}, 
%\begin{subequations}
\begin{align}
\label{eq9}
|\psi_{00}\rangle 
&= \frac{1}{\sqrt{3}}\left( |1_c 1_t\rangle + |2_c 2_t\rangle + |3_c 3_t\rangle \right), \nonumber \\[1em]
|\psi_{12}^+\rangle 
&= \frac{1}{\sqrt{2}}\left( |1_c 2_t\rangle + |2_c 1_t\rangle \right), 
\nonumber \\[1em]
|\psi_{12}^-\rangle 
&= \frac{1}{\sqrt{2}}\left( -|1_c 2_t\rangle + |2_c 1_t\rangle \right), \nonumber
 \\[1em]
|\psi_{13}^+\rangle 
&= \frac{1}{\sqrt{2}}\left( |1_c 3_t\rangle + |3_c 1_t\rangle \right), 
\nonumber \\[1em]
|\psi_{13}^-\rangle
&= \frac{1}{\sqrt{2}}\left( -|1_c 3_t\rangle + |3_c 1_t\rangle \right), \\[1em]
%\label{eq9}
%\end{align}
%\vspace{1em}
%\begin{align}
|\psi_{23}^+\rangle 
&= \frac{1}{\sqrt{2}}\left( |2_c 3_t\rangle + |3_c 2_t\rangle \right), 
\nonumber \\[1em]
|\psi_{23}^-\rangle 
&= \frac{1}{\sqrt{2}}\left( -|2_c 3_t\rangle + |3_c 2_t\rangle \right),  \nonumber \nonumber\\[1em]
|\psi_{33}\rangle 
&= \frac{1}{\sqrt{2}}\left( -|2_c 2_t\rangle + |3_c 3_t\rangle \right), 
\nonumber \\[1em]
|\psi_{88}\rangle 
&= \frac{1}{\sqrt{6}}\left(- 2|1_c 1_t\rangle + |2_c 2_t\rangle + |3_c 3_t\rangle \right). \nonumber 
\end{align}
For the $VT$ system, using the interaction Hamiltonian in Eq.~(\ref{eq4}), the transition matrix elements between the separable state $|1_c,1_t\rangle$ and the $SU(3)$ entangled states in Eq.~(\ref{eq9}) are given by, 
\begin{subequations}
\label{eq10}
\begin{align}
\langle 1_c 1_t \vert H^\text{VT}_{\text{I,c-t}} \vert \psi_{ab} \rangle
&= 0, \quad \forall \, \psi_{ab} \in \{\psi_{00}, \psi_{23}^\pm, \psi_{33},\psi_{88}\},
\label{eq10a}\\
\langle 1_c 1_t \vert H^\text{VT}_{\text{I,c-t}} \vert \psi_{12}^+ \rangle
&= \frac{1}{2\sqrt{2}} \bigl( \Omega_{12}^{(c)} + \Omega_{12}^{(t)} \bigr),
\label{eq10b}\\
\langle 1_c 1_t \vert H^\text{VT}_{\text{I,c-t}} \vert \psi_{12}^- \rangle
&= \frac{1}{2\sqrt{2}} \bigl( \Omega_{12}^{(c)} - \Omega_{12}^{(t)} \bigr),
\label{eq10c}\\
\langle 1_c 1_t \vert H^\text{VT}_{\text{I,c-t}} \vert \psi_{13}^+ \rangle
&= \frac{1}{2\sqrt{2}} \bigl( \Omega_{13}^{(c)} + \Omega_{13}^{(t)} \bigr),
\label{eq10d}.\\
\langle 1_c 1_t \vert H^\text{VT}_{\text{I,c-t}} \vert \psi_{13}^- \rangle
&= \frac{1}{2\sqrt{2}} \bigl( \Omega_{13}^{(c)} - \Omega_{13}^{(t)} \bigr)
\label{eq10e}
\end{align}
\end{subequations}
\noindent 
These matrix elements reveal that only the entangled states $\{|\psi_{12}^{\pm}\rangle, |\psi_{13}^{\pm}\rangle\}$  remain dynamically coupled to the intermediary state, while the remaining states are completely decoupled due to the underlying symmetry. The treatment for the $TU$ and $UV$ systems is similar and in Table I we have displayed the expectation values of all three systems: 
\begin{table}[h]
\caption{\label{tab:coupling_strengths} Expectation value $\langle i_c i_t | H^{\text{MN}}_{\text{I,c-t}} | \psi_{ab} \rangle$ for $VT$, $TU$, and $UV$ system}
\begin{ruledtabular}
\begin{tabular}{lccc}
State $|\psi_{ab}\rangle$ & $\langle 1_c 1_t | H_{\text{I,c-t}}^\mathrm{VT} | \psi_{ij} \rangle$ & $\langle 2_c 2_t | H_{\text{I,c-t}}^\mathrm{TU} | \psi_{ij} \rangle$ & $\langle 3_c 3_t | H_{\text{I,c-t}}^\mathrm{UV} | \psi_{ij} \rangle$ \\
\colrule
%$|\psi_{00}\rangle$ & $0$ & $0$ & $0$ \\
$|\psi_{12}^+\rangle$ & $\frac{\Omega_{12}^{(c)} + \Omega_{12}^{(t)}}{2\sqrt{2}}$ & $\frac{\Omega_{12}^{(c)} + \Omega_{12}^{(t)}}{2\sqrt{2}}$ & $0$ \\
$|\psi_{12}^-\rangle$ & $\frac{\Omega_{12}^{(c)} - \Omega_{12}^{(t)}}{2\sqrt{2}}$ & $\frac{\Omega_{12}^{(c)} - \Omega_{12}^{(t)}}{2\sqrt{2}}$ & $0$ \\
$|\psi_{13}^+\rangle$ & $\frac{\Omega_{13}^{(c)} + \Omega_{13}^{(t)}}{2\sqrt{2}}$ & $0$ & $\frac{\Omega_{13}^{(c)} + \Omega_{13}^{(t)}}{2\sqrt{2}}$ \\
$|\psi_{13}^-\rangle$ & $\frac{\Omega_{13}^{(c)} - \Omega_{13}^{(t)}}{2\sqrt{2}}$ & $0$ & $\frac{\Omega_{13}^{(c)} - \Omega_{13}^{(t)}}{2\sqrt{2}}$ \\
$|\psi_{23}^+\rangle$ & $0$ & $\frac{\Omega_{23}^{(c)} + \Omega_{23}^{(t)}}{2\sqrt{2}}$ & $\frac{\Omega_{23}^{(c)} + \Omega_{23}^{(t)}}{2\sqrt{2}}$ \\
$|\psi_{23}^-\rangle$ & $0$ & $\frac{\Omega_{23}^{(c)} - \Omega_{23}^{(t)}}{2\sqrt{2}}$ & $\frac{\Omega_{23}^{(c)} - \Omega_{23}^{(t)}}{2\sqrt{2}}$ \\
%$|\psi_{33}\rangle$ & $0$ & $0$ & $0$ \\
$|\psi_{00}\rangle$, $|\psi_{33}\rangle$, $|\psi_{88}\rangle$ & $0$ & $0$ & $0$ 
\end{tabular}
\end{ruledtabular}
\end{table}

Having identified the bright states, we proceed to define the synthetic basis manifold
$\mathcal{M}_{\rm{dyn}}^{\rm MN}=
{\rm span}\Bigl\{
\ket{g_c g_t}, \ket{\psi_{ij}^+} :
\langle g_c g_t \vert H_{\rm I,c-t}^{\rm MN} \vert \psi_{ij}^+ \rangle \neq 0
\Bigr\}$,
which contains the collective states actively participating in the dynamics, while the remaining antisymmetric states $\ket{\psi_{ij}^-}$ become fully dark. For example, in the $VT$ configuration under site-independent driving $\Omega_{ij}^{(c)}=\Omega_{ij}^{(t)}\equiv \Omega_{ij}$,
the antisymmetric states Eqs.~(\ref{eq10c}) and (\ref{eq10e})] are completely decoupled from the dynamics, leaving only the contributing transition amplitudes
\begin{align}
\langle 1_c 1_t|H^\text{VT}_{\mathrm{I,c\text{-}t}}|\psi_{12}^+\rangle = \frac{\Omega_{12}}{\sqrt{2}}, \quad 
\langle 1_c 1_t|H^\text{VT}_{\mathrm{I,c\text{-}t}}|\psi_{13}^+\rangle = \frac{\Omega_{13}}{\sqrt{2}}.
\label{eq11}
\end{align}
Another critical observation arises from the interaction matrix element where the transition amplitude between the two symmetric entangled states vanishes, 
\begin{align}
\langle{\psi_{12}^+}| H^{\text{VT}}_{\text{I,c-t}} |\psi_{13}^+\rangle = 0.
\label{eq12}
\end{align}
Thus Eq.~(\ref{eq11}), together with Eq.~(\ref{eq12}) yield a selection rule of the dynamics which defines a characteristic  structured $VT$  configuration with the reduced manifold containing the synthetic basis states,  
\begin{align} 
\mathcal{M}_{\rm{dyn}}^{\text{VT}} = {\rm span}\{|1_c, 1_t\rangle, |\psi_{12}^+\rangle, |\psi_{13}^+\rangle\}, 
\label{eq13}
\end{align}
with corresponding effective Hamiltonian given by, 
\begin{align}
\begin{aligned}
\text{H}^\text{VT} &= \frac{1}{3} E_{11}^\text{{VT}} |1_c,1_t\rangle\langle 1_c,1_t| + \frac{1}{3} E_{22}^\text{{VT}} |\psi_{12}^+\rangle\langle \psi_{12}^+| + \frac{1}{3} E_{33}^\text{VT} |\psi_{13}^+\rangle\langle \psi_{13}^+|\\ 
&+  \frac{\Omega_{12}}{\sqrt{2}} |1_c,1_t\rangle\langle \psi_{12}^+| + \frac{\Omega_{13}}{\sqrt{2}} |1_c,1_t\rangle\langle \psi_{13}^+| + \text{h.c.}, 
\end{aligned}
\label{eq14}
\end{align}
where $\Omega_{ij}$ represents the effective Rabi frequency characterizing the collective coupling within the $VT$ system.
\par 
Proceeding in the similar way, for the $TU$ configuration we have following non-vanishing transitions with the intermediary state $|2_c 2_t\rangle$, 
\begin{align}
\langle 2_c 2_t | H^\text{TU}_\text{I,c-t} | \psi_{12}^+ \rangle = \frac{1}{\sqrt{2}}\Omega_{12}, \quad 
\langle 2_c 2_t | H^\text{TU}_\text{I,c-t} | \psi_{23}^+  \rangle = \frac{1}{\sqrt{2}}\Omega_{23},
\label{eq15}
\end{align}
along with the absence of direct coupling between the two symmetric states, $\langle\psi_{12}^+|H^{\text{TU}}_{\text{I,c-t}}|\psi_{23}^+\rangle = 0$. This configuration ensures that the entanglement dynamics is strictly confined within reduced manifold $\mathcal{M}_{\rm{dyn}}^{\text{TU}} = {\rm span}\{|\psi_{12}^+\rangle, |2_c 2_t\rangle, |\psi_{23}^+\rangle\}$
and the Hamiltonian in this basis is given by,
\begin{align}
\begin{aligned}
\text{H}^\text{TU} &= \frac{1}{3} E_{11}^\text{TU}|\psi_{12}^+\rangle \langle \psi_{12}^+| + \frac{1}{3} E_{22}^\text{TU} |2_c 2_t\rangle\langle 2_c 2_t| + \frac{1}{3} E_{33}^\text{TU} |\psi_{23}^+ \rangle \langle \psi_{23}^+| \\ 
&+  \frac{\Omega_{12}}{\sqrt{2}} |2_c 2_t\rangle \langle \psi_{12}^+| + \frac{\Omega_{23}}{\sqrt{2}} |2_c 2_t\rangle\langle \psi_{23}^+| + \text{h.c.},
\end{aligned}
\label{eq16}
\end{align}
Similarly for the $UV$ configuration with the state $|3_c 3_t\rangle$ becomes intermediate state with following transition matrix are $\langle 3_c 3_t | H^\text{UV}_\text{I,c-t} | \psi_{23}^+ \rangle = \frac{\Omega_{23}}{\sqrt{2}}$, $\langle 3_c 3_t | H^\text{UV}_\text{I,c-t} | \psi_{13}^+ \rangle = \frac{\Omega_{13}}{\sqrt{2}}$ and $\langle\psi_{13}^+|H^{\text{UV}}_{\text{I,c-t}}|\psi_{23}^+\rangle = 0$, 
the Hamiltonian in this basis is given by,
\begin{align}
\begin{aligned}
H^\text{UV} &= \frac{1}{3} E_{11}^\text{UV} |\psi_{13}^+\rangle \langle \psi_{13}^+| + \frac{1}{3} E_{22}^\text{UV} |\psi_{23}^+\rangle\langle \psi_{23}^+| + \frac{1}{3} E_{33}^\text{UV} |3_c 3_t \rangle \langle 3_c 3_t| \\
&\quad+ \frac{\Omega_{13}}{\sqrt{2}} |3_c 3_t\rangle \langle\psi_{13}^+| + \frac{\Omega_{23}}{\sqrt{2}} |3_c 3_t\rangle\langle \psi_{23}^+| + \text{h.c.}, 
\end{aligned}
\label{eq17}
\end{align}
with reduced manifold $\mathcal{M}_{\rm{dyn}}^{\text{UV}} = {\rm span}\{|\psi_{13}^+\rangle, |\psi_{23}^+\rangle,|3_c3_t\rangle\}$. Thus, all three synthetic three-level systems emerge entirely from collective $SU(3)$ entangled modes embedded within the composite Hilbert space, rather than from bare atomic energy level.
\par 
A fundamental distinction between the present synthetic three-level architecture and conventional bare atomic three-level systems should be emphasized. In ordinary atomic systems, the energy levels correspond to bare atomic states that are typically treated on an equal footing under permutation symmetry. In contrast, the synthetic configurations considered here emerge from a composite two-qutrit manifold consisting of a separable product state, which acts as an intermediary coupling state, together with two collective entangled states generated by the underlying $SU(3)$ symmetry. The specific choice of the reference product state explicitly breaks the permutation symmetry of the effective manifold and thereby introduces distinguishability among the synthetic levels. 
In particular, the states $\ket{1_c1_t}$, $\ket{2_c2_t}$, and $\ket{3_c3_t}$ serve as the common coupling centers for the synthetic $VT$, $TU$, and $UV$ configurations, respectively. A schematic illustration of the corresponding mappings onto synthetic $V$-, $\Xi$-, and $\Lambda$-type three-level configurations is shown in Fig.~1. This symmetry‑broken collective structure necessitates a new framework for revisiting a broad class of coherent quantum‑optical phenomena, particularly those arising in engineered three‑level with two qutrits. 

\section{Measure for synthetic three-level system} 

Finally, to quantify entanglement for the synthetic three-level systems build-up from two-qutrit configuration, we consider the I-concurrence \cite{Rungta2001}, which naturally extends to higher-dimensional Hilbert spaces, together with a generalized variant of Wootters-type concurrence formulated within the $SU(3)$ framework \cite{Wootters1998}.

\subsection{$SU(3)$ I-Concurrence}

The I-concurrence for a high-dimensional entangled state can be expressed in terms of the purity of the reduced subsystem as
\cite{Rungta2001},
\begin{align}
C_{I}(\ket{\psi_{ij}}) = \sqrt{2\left(1 - \operatorname{Tr}[\rho_c^2]\right)},
\label{eq18}
\end{align}
where $\rho_c(t)$ (or $\rho_t(t)$) is the reduced density matrix of the control (or target) subsystem for the given state. To analyze the dynamical evolution of the three configurations, we first consider the time-dependent wave function of the $  VT  $ configuration governed by the Hamiltonian in Eq.~(\ref{eq4}). The corresponding time-dependent state vector is given by 
\begin{align}
\ket{\Psi^{VT}(t)} = \alpha(t) \ket{1_c1_t} + \beta(t) \ket{\psi_{12}^+} + \gamma(t) \ket{\psi_{13}^+},
\label{eq19}
\end{align}
where $\alpha(t)$, $\beta(t)$ and $\gamma(t)$ be the normalized amplitudes. Plucking back the symmetric basis states $SU(3)$ entangled states $\ket{\psi_{13}^+}$ and $\ket{\psi_{12}^+}$ from Eq.~(\ref{eq9}) in Eq.~(\ref{eq19}) we have, 
%Eq.(21) can be written as,
\begin{align}
\ket{\Psi^{VT}(t)} =
\alpha(t)\ket{1_c1_t}
+ \frac{\beta(t)}{\sqrt{2}}(\ket{1_c3_t}+\ket{3_c1_t})
+ \frac{\gamma(t)}{\sqrt{2}}(\ket{1_c2_t}+\ket{2_c1_t}).
\label{eq20}
\end{align}
Taking partial trace, the reduced density matrix of the control (or target) subspace obtained from Eq.(\ref{eq20}) is given by,
\begin{align}
\rho_c^{VT}(t) = 
\begin{pmatrix}
|\alpha(t)|^2 + \frac{|\beta(t)|^2}{2} + \frac{|\gamma(t)|^2}{2} & \frac{\alpha(t) \beta^*(t)}{\sqrt{2}} & \frac{\alpha(t)\gamma^*(t)}{\sqrt{2}} \\
\frac{\beta(t) \alpha^*(t)}{\sqrt{2}} & \frac{|\beta(t)|^2}{2} & \frac{\beta(t) \gamma(t)^*}{2} \\
\frac{\gamma(t) \alpha(t)^*}{\sqrt{2}} & \frac{\gamma(t) \beta(t)^*}{2} & \frac{|\gamma(t)|^2}{2}
\end{pmatrix}, 
\label{eq21}
\end{align}  
and plugging it back into Eq.(\ref{eq18}) gives the $SU(3)$ I-Concurrence of the $VT$ system,
\begin{align}
C_I^{VT}(t) = |\beta(t)|^2 + |\gamma(t)|^2, 
\label{eq22}
\end{align}
confirming that the I-concurrence can be used as a robust metric for tracking the coherent dynamics of the synthetic three-level system which depends upon the population residing in the symmetric excited states. This, in addition, reveals that the effective manifold $\mathcal{M}_{\rm dyn}^{\rm VT}$ can be decomposed into a direct sum of separable and entangled sectors, i.e., 
$\mathcal{M}_{\rm dyn}^{\rm VT} = \mathcal{M}_{\rm 1_c1_t} \oplus \mathcal{M}_{\rm excited}$. 

Proceeding in a similar manner, the wave functions corresponding to the synthetic $TU$ and $UV$ configurations may be written as, 
\begin{align}
\ket{\Psi^{\text{TU}}(t)}=
\alpha(t)\ket{\psi_{12}^+} +\beta(t)\ket{2_c 2_t}
+\gamma(t)\ket{\psi_{23}^+},
\label{eq23}
\end{align}
and
\begin{align}
\ket{\Psi^\text{UV}(t)}=
\alpha(t)|\psi_{23}^+\rangle+\beta(t)|\psi_{13}^+\rangle+
\gamma(t)|3_c3_t\rangle,
\label{eq24}
\end{align}
respectively and following the same procedure the corresponding reduced density matrices and $SU(3)$ I-concurrences can be obtained. Table~II summarizes the I-concurrence for various $SU(3)$ entangled states (upper panel), satisfying the bound $1 \leq C_I(\psi_{ij}) \leq \frac{2}{\sqrt{3}}$, together with the three synthetic configurations (lower panel): 
\begin{table}[h]
\caption{\label{tab:Table2} Comparison of I-Concurrence and $SU(3)$ Concurrence}
\begin{ruledtabular}
\begin{tabular}{ccc}
State & $C_{I}(\ket{\psi_{ij}})$ & $C_{SU(3)}(\ket{\psi_{ij}})$ \\
\hline
\hline
 & $SU(3)$ Entangled State ($\ket{\psi_{ij}}$) & \\ 
\hline
$\ket{\psi_{00}}$ & $2/\sqrt{3}=1.155$ & 2 \\
$\ket{\psi_{12}^+}, \ket{\psi_{13}^+}, \ket{\psi_{23}^+}$ & 1 & 1 \\
$\ket{\psi_{12}^-}, \ket{\psi_{13}^-}, \ket{\psi_{23}^-}$ & 1 & 1 \\
$\ket{\psi_{33}}, \ket{\psi_{88}}$ & 1 & 1\\
\hline
& Synthetic Three-level System ($\ket{\Psi^\mathrm{MN}}$) & \\ 
\hline
$\ket{\Psi^{VT}}$ & $|\beta(t)|^2 + |\gamma(t)|^2$ & $|\beta(t)|^2 + |\gamma(t)|^2$ \\
$\ket{\Psi^{TU}}$ & $|\alpha(t)|^2 + |\gamma(t)|^2$ & $|\alpha(t)|^2 + |\gamma(t)|^2$ \\
$\ket{\Psi^{UV}}$ & $|\alpha(t)|^2 + |\beta(t)|^2$ & $|\alpha(t)|^2 + |\beta(t)|^2$ \\
\end{tabular}
\end{ruledtabular}
\end{table}

\subsection{SU(3) Concurrence}
\par 
Another important measure of entanglement is the Wootters concurrence \cite{Wootters1998}, originally formulated for two-qubit systems through a spin-flip transformation of the $SU(2)$ group. While several extensions to higher dimensions have been proposed \cite{Albeverio2001}, a simple and physically transparent formulation suitable for synthetic three-level qutrit systems has been lacking. In this work, we utilize the antisymmetric Gell-Mann operators of the $SU(3)$ algebra to obtain a natural generalization of Wootters' concurrence for two-qutrit systems. 

To evaluate the generalized $SU(3)$ concurrence, we consider a general two-qutrit state in terms of the computational basis  used to define the $SU(3)$ entangled states in Eq.(\ref{eq9}), 
\begin{align}
|\psi\rangle = \sum_{i,j=1}^{3} c_{ij}|ij\rangle,
\label{eq25}
\end{align}
%where the $SU(3)$ entangled states introduced in Eq.~(9)
and corresponding dual (spin-flipped) state is given by,
\begin{align}
|\tilde{\psi}\rangle =
\Big[(\lambda_2 \otimes \lambda_2) + (\lambda_5 \otimes \lambda_5) + (\lambda_7 \otimes \lambda_7)\Big] |\psi^*\rangle,
\label{eq26}
\end{align}
where $\lambda_2$, $\lambda_5$, and $\lambda_7$ are the antisymmetric Gell-Mann matrices generating pairwise transitions within the three embedded $SU(2)$ subspaces associated with the level pairs (1,2), (1,3), and (2,3) of the qutrit system. The $SU(3)$ concurrence is defined through the overlap between the state and its dual counterpart as \cite{Wootters1998}
\begin{align}
C_{SU(3)} = \big|\langle \psi \mid \tilde{\psi} \rangle \big|.
\label{eq27}
\end{align}
Evaluating this overlap yields a compact expression in terms of the $2\times2$ principal minors of the coefficient matrix \cite{Albeverio2001}, 
\begin{align}
C_{SU(3)}(\psi_{ij}) = \sqrt{2 \sum_{i < j} \sum_{k < l} |c_{ik} c_{jl} - c_{il} c_{jk}|^2}
\label{eq28}
\end{align}

For the synthetic $VT$ configuration described by the wave function in Eq.~(\ref{eq19}), the non-vanishing coefficients are $c_{11}=\alpha(t)$, $  c_{13}=c_{31}=\frac{\beta(t)}{\sqrt{2}}$, and $  c_{12}=c_{21}=\frac{\gamma(t)}{\sqrt{2}}$. Consequently, only the minors associated with the states $|\psi_{12}^+\rangle$ and $|\psi_{13}^+\rangle$ contribute, 
and the $SU(3)$ concurrence in Eq.~(\ref{eq28}) is reduced to,
\begin{align}
C_{SU(3)}^{VT}(t) = |\beta(t)|^2 + |\gamma(t)|^2.
\label{eq29}
\end{align}
It is worth noting that a direct comparison between the $SU(3)$ I-concurrence obtained from the reduced density matrix in Eq.~(\ref{eq22}) and aforesaid generalized $SU(3)$ spin-flip concurrence shows that both measures yield identical result,
\begin{align}
C_{\mathrm{SU}(3)}^{VT}(t)=C_{I}^{VT}(t),
\label{eq30}
\end{align}
while the separable reference state $\ket{1_c1_t}$ does not contribute to the entanglement measure. An analogous result follows for the remaining $TU$ and $UV$ configurations through cyclic permutation of the basis states, where the separable states $\ket{2_c2_t}$ and $\ket{3_c3_t}$, respectively, remain noncontributing. Table~II summarizes the $SU(3)$ concurrence for the three synthetic configurations together with the entangled states defined in Eq.~(\ref{eq9}), including the maximally entangled state $\ket{\psi_{00}}$, satisfying the bound $1 \leq C_{\mathrm{SU}(3)} \leq 2$.

\section{Conclusion}

In this work, we have developed a two-qutrit architecture based on two entangled three-level subsystems. By exploiting the algebraic structure of the $SU(3)$ group in composite Hilbert space together with the recently introduced $SU(3)$ entangled states, we identify synthetic manifold  consists of two entangled states along with a separable intermediary state. These collective modes emulate the effective Hamiltonian of all synthetic three-level configurations without introducing Rydberg state or van der Waals interaction. We further formulate the $SU(3)$ I-concurrence and a generalized Wootters-type concurrence, both of which provide a consistent and physically transparent measures of the nonlocal correlations present in these synthetic systems. In summary, the present work establishes a unified $SU(3)$-based framework for the realization of synthetic three-level systems, thereby opening a promising route toward high-dimensional quantum information processing beyond conventional Rydberg-mediated qubit architectures.
\vfill 

\pagebreak 

%\newpage 

\begin{figure*}[t]
\centering
\hspace{0.02\textwidth}
\includegraphics[width=0.8\textwidth]{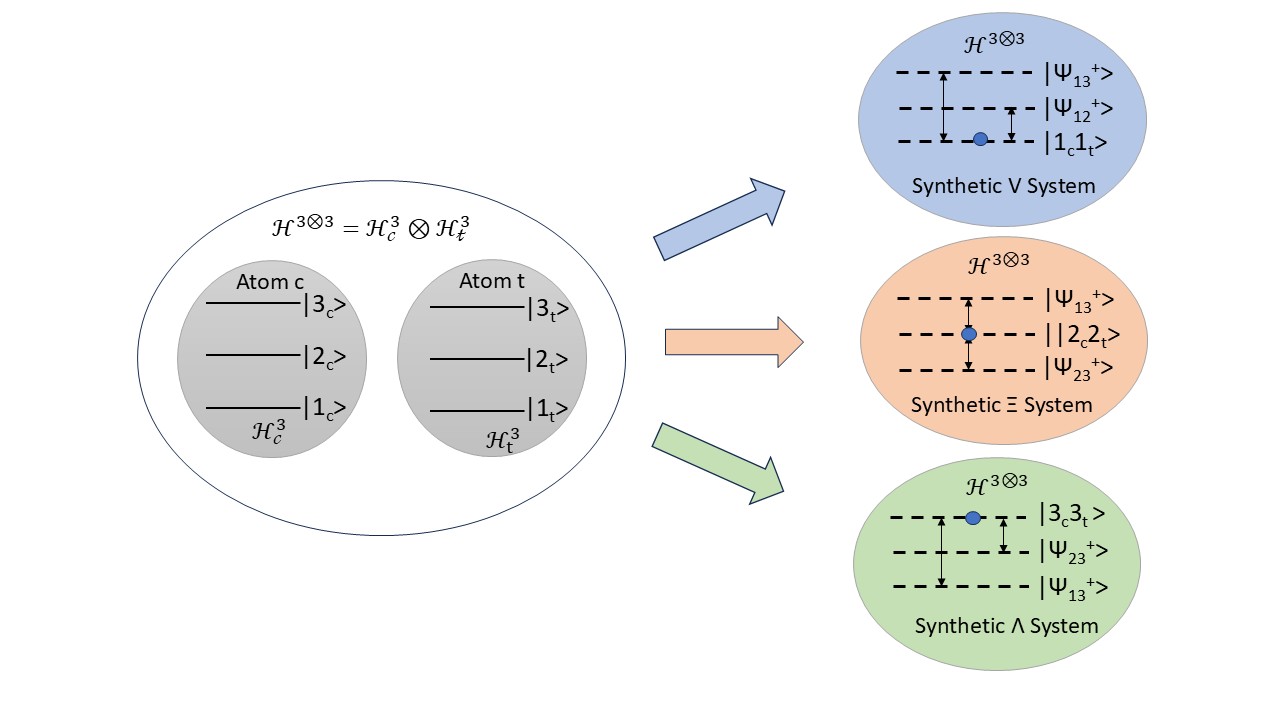}
\caption{Schematic illustration of the mapping of a two-qutrit system onto synthetic $V$, $\Xi$ and $\Lambda$ type three-level configurations. The separable reference state \( |i_c i_t\rangle \) is represented by a dot, which serves as the common intermediary state, ensures the distinguishability of the energy levels within each synthetic hierarchy.}
\label{fig:configurations}
\end{figure*}
%END%%%%%%%%%%%%%%%%%%%%%%%%%%%%%%%%%%%%%%%%%%%%%%%%

\pagebreak 

%\newpage 

\bibliography{references}% Produces the bibliography via BibTeX.

\end{document}